\documentclass[prb]{revtex4}
\usepackage{graphicx}
\begin{document}

\title{Geometric Aspects of the Dipolar Interaction in Lattices of Small Particles}
\author{Paola R. Arias, D. Altbir}
\affiliation{Departamento  de F\'{\i}sica, Universidad de
Santiago de  Chile, USACH\\ Avenida Ecuador 3493, Santiago, Chile}
\author{M. Bahiana} 
\affiliation{Instituto de F\'{\i}sica, Universidade Federal do Rio de Janeiro\\
Caixa Postal 68528, Rio de Janeiro, RJ, Brazil, 21945-972}
\begin{abstract} 
The hysteresis curves of systems composed of small interacting magnetic particles,
regularly placed on stacked layers, are obtained with Monte Carlo simulations. The remanence as a function of temperature, in interacting systems, presents a peak that separates two different magnetic states. At low temperatures, small values of remanence are a consequence of antiferromagnetic order due to the dipolar
interaction. At higher values of temperature the increase of the component  normal to the lattice plane is responsible for the  
small values of remanence. 
The effect of the number of layers, coordination number and distance between particles are investigated.
\end{abstract}
\pacs{75.10.-b, 75.20.-g,75.75.+a} 
\maketitle
\section{Introduction} \label{sec:Intro}
In the past decade the magnetism of fine particles embedded in a non magnetic
matrix has been a topic of interest because of their uses on chemical catalysis and magnetic recording \cite{recording}. The appearance of new experimental techniques capable of generating samples with controlled nanostructures\cite{luis1,dots} has led to
important advances in the preparation and understanding of the behavior of granular materials.  However, at the nanometer scale, magnetic systems are not easily reproduced and characterized, introducing  difficulties for the investigation of these systems. 
Experimental and theoretical results over the past years show that there are clearly many factors that influence the magnetic and magnetotransport behavior of these systems, such as the distribution of grain sizes, the average size and shape of the grains
and the magnetic anisotropy of the individual grains. Also  the role of magnetic interactions among crystallites is a topic full of controversies, despite the intense research on the subject. One of the most used methods to investigate the role of interactions has been Monte Carlo simulations.  El-Hilo {\it et al.}\cite{el-hilo} had used it for determining the magnetoresistance dependence on the mean intergranular distance, or rather, the particle concentration, using a simple expression previously obtained by Gittleman {\it et al.}\cite{gittleman} Also Garcia-Otero {\it et al.}\cite{garcia1} analyzed the interplay between anisotropy and magnetic interactions and Chantrell {\it et al.}\cite{chantrell} calculated the susceptibility and ZFC-FC (Zero Field Cooled-Field Cooled) magnetization curves for superparamagnetic particles. Some simple models taking the dipolar interaction into account have been proposed. For example,  M\o rup and Tronc \cite{morup1} have formulated a description for weakly interacting particles, and Dormann {\it et al.}\cite{dormann2} have proposed the Dormann-Bessais-Fiorani (DBF) model, valid for weak and medium strength of the interactions. 
Both models lead to contradictory results, as analyzed in Ref. [10]. The discrepancy arises when we try to determine if increasing concentration, the interactions leads to increase or decrease the energy barrier of the system. More recently, Allia {\it et al.}\cite{allia} have proposed analytical models that take explicitly into account the correlation arising from the dipolar interactions on nearly superparamagnetic systems and Pike {\it et al.}\cite{Pike} investigated the role of magnetic interactions on low temperature  saturation remanence of fine magnetic particles. 
From the experimental point of view, while dilute systems are well understood, results for denser ones, where the interactions between particles play an important role, are still not clear. The main reasons are the unavoidable particle size distribution and difficulties in controlling and replicating geometrical arrangement and orientation of easy axis for higher concentrations. In particular we can mention that for Fe particles embedded in an alumina matrix\cite{sahoo} and for $\gamma-Fe_2O_3$\cite{exp2} particles, an increase of the blocking temperature $T_B$ with interaction strength is obtained. However, also for $\gamma-Fe_2O_3$ particles investigated by Mossbauer spectroscopy $T_B$ decreases with increasing concentration, as presented in [14]. Apart from  the influence of concentration, or interparticle distance, in systems for which the dipolar interaction may not be neglected, it is important to consider the effect of dimensionality, specially when one deals with systems formed from sequential deposition of layers. \cite{luis1,sahoo,comluis1} Luis {\it et al.}\cite{luis1} investigated the role of dipolar interactions on the magnetization of Co clusters grown in a quasi-ordered layered structure, and  showed that the affective activation energy increases linearly with the number of nearest neighbor clusters. These results have been interpreted in terms of a transition from 2D to 3D collective in Ref. [15]. Bahiana {\it et al.}\cite{ordering} investigated the effect of interactions on a 2D lattice of grains and found a peak on the remanence as a function of temperature, which can be interpreted as an evidence of the existence of two low magnetization states: one due to an in-plane alignment perpendicular to the field direction at low temperatures, and a high temperature disordered state. In this paper we have extended these results to 3D systems, considering the effect of increasing the number of stacked layers, and lattice geometry. The hysteresis curves  are simulated by means of Monte Carlo method. 

\section{Simulation conditions}
A ferromagnetic particle becomes a monodomain when its linear size is below a critical value $D_c$ determined by the minimization of the total energy, including magnetostatic, exchange and anisotropy contributions.\cite{bertotti}  Such monodomain ferromagnetic particles can be viewed as large magnetic units, each one having a magnetic moment of thousands of Bohr magnetons.  For neighboring particles separated by $10 - 30$ nm, direct and indirect exchange (like RKKY) can be neglected \cite{altbir1},thus, the magnetic properties of such assembly of nanoparticles are determined by the dipolar interaction energy among the particles along with thermal and magnetic anisotropy energies. The magnetic irreversibility of an isolated nanomagnet is conventionally associated to the energy required for the particle moment reorientation, overcoming a barrier due to shape, magnetoelasticity, and/or crystalline anisotropy.\cite{bertotti} In the presence of relevant dipolar interactions, this simplified picture no longer holds, as each particle is subject to a complicated energy landscape.

\begin{figure}
\begin{center}
\includegraphics{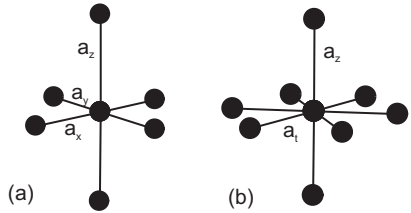}
\caption{Different geometries considered in the simulations. (a) $xy$ planes with rectangular arrangement. In this case the grains are separated by $a_x$ and $a_y$ along the $x$ and $y$ axis respectively, separated by $a_z$ on the $z$ direction.(b) $xy$ planes with a triangular symmetry and nearest neighbor distance $a_t$ are separated by a distance $a_z$. \label{fig-redes}}
\end{center}
\end{figure}

To investigate the magnetic behavior of interacting grains we have examined two simple systems consisting of $M$ layers of $N \times N$ magnetic 3D monodomain particles. Each layer is parallel to the $xy$ plane and has free boundary conditions. Two nearest neighbors arrangements were considered, with square and triangular symmetries, as shown in Fig.~\ref{fig-redes}. The distance between particles on each layer is defined by the lattice parameters $a_x$ and $a_y$, for square lattices, and $a_t$ for triangular lattices, the layers separation is given by $a_z$. From now on the terms in-plane and out-plane refer to the $xy$ plane.
The particles have uniform magnetization, $m=869\mu_B$, and anisotropy constant, $K=1.32 \times 10^6$ erg/cm$^3$, corresponding to slightly elongated Cobalt grains with about 511 atoms and linear dimensions of the order of 20 \AA.\cite{cullity} Each particle is described by the position of its center of mass and the direction of the randomly chosen 3D easy magnetization axis, $\widehat{e}_{i}$, and are coupled by means of dipolar interactions. Since we seek to understand the role of lattice geometry, we prefer to control the distance between grains through the parameters $a_x$, $a_y$, $a_z$ and $a_t$, instead of concentration. The values of lattices parameters are such that the particles surfaces are more than one Cobalt lattice parameter apart. \cite{ordering} The external field is always in the $x$ direction.

In the presence of an external magnetic field $\overrightarrow{H}$, the total energy of the system is written as
\begin{equation}
\mathcal{E}=\sum_{i}\left[-\overrightarrow{m}_{i}\cdot \overrightarrow{H}-
\kappa \left( \frac{\overrightarrow{m}_{i}\cdot \widehat{e}_{i}}{m_{i}}\right)^{2}+\frac{1}{2}\sum_{j\neq i} E_{ij} \right] \; \;,
\label{hamiltonian}
\end{equation}
where $\kappa = K V$, $V$ being the volume of each grain. $E_{ij}$ is the classical dipolar energy between grains $i$ and $j$ given by
\[
E_{ij}=\frac{\overrightarrow{m }_{i}\cdot \overrightarrow{m }_{j}-3(
\overrightarrow{m }_{i}\cdot \widehat{n}_{ij})(\overrightarrow{m }
_{j}\cdot \widehat{n}_{ij})}{r_{ij}^{3}}\;\;.
\]
Here $r_{ij}$ is the distance between the centers of particles ${i}$ and $j$, and $\widehat{n}_{ij}$ is the unit vector along the direction that connects them.
Using this expression for the energy we have simulated hysteresis curves for different values of temperature and lattice symmetries. Hysteresis curves correspond  to sequences of nonequilibrium states of the system, and therefore depend on the field variation rate. In terms of a Monte Carlo simulation this means that we have to avoid equilibrium by a sufficiently fast variation of the field, and the usual mechanism of time averaging instead of ensemble averaging is not valid. Monte Carlo
simulations were carried out using Metropolis algorithm with local dynamics in which the new orientation of the magnetic moment was chosen within a solid angle around the previous moment direction, with aperture $d\theta=d\phi=0.1$. This method was studied by Pereira Nunes {\it et al.}\cite{neqmc}, applied to the simulation of ZFC-FC magnetization curves, which are also a collection of nonequilibrium states, and it is a reasonable approach for a qualitative analysis.
Monte Carlo simulations of hysteresis curves have been used in a variety of magnetic systems with good qualitative agreement with experimental data, so we believe that, although not specifically created to describe nonequilibrium processes, the method provides a valuable tool for studying the dynamics of complex systems.\cite{sampaio,mporto2,chantrell} For simulating hysteresis, we started at a fixed temperature from a configuration in which the magnetic moment directions were randomly chosen. An external magnetic
field $H=0.25$ kOe in the $x$ direction was turned on, one of the grains was randomly chosen, and had its magnetic moment rotated by an angle restricted to a cone, as explained above.  The change in energy ($\Delta\mathcal{E}$) was calculated and the rotation accepted with probability $p=\mbox{min}[1,\exp(-\Delta \mathcal{E}/k_B T)]$.  This procedure was repeated $N^2\times M$ times, comprising one Monte Carlo step. The number of
Monte Carlo steps in nonequilibrium simulations is a rather arbitrary choice, as explained by Pereira Nunes {\it et al.} \cite{neqmc} Actually, the variation rate of the external field is the important quantity.  In this case, we first fixed the value of the variation step for the external field, $dH$, with the objective of having enough points in
the region of interest, near  $H=0$, but still being able to bring the system to saturation. We found
that  200 Monte Carlo steps was a good choice for $dH=0.25\;$kOe, since the system shows hysteresis for some values of temperature, and the area of the hysteresis cycle shows sensitivity to temperature variation.
The virgin curve was then obtained by increasing the field until the magnetization reached at least 99.995\% of its saturation value, $m_s$. Starting from this last value, the field was decreased to negative
values at the same rate (200 Monte Carlo steps per $dH=0.25\;$kOe). The whole procedure was repeated 5 to 20 times, depending on the system size,  for different random choices of easy magnetization axis directions, the averaged hysteresis curve was calculated, and the remanence, $m_r$, was determined as the $x$ component of the magnetization for $H=0$.

\section{Results}
\subsection{Square lattices}
\begin{figure}
\begin{center}
\includegraphics{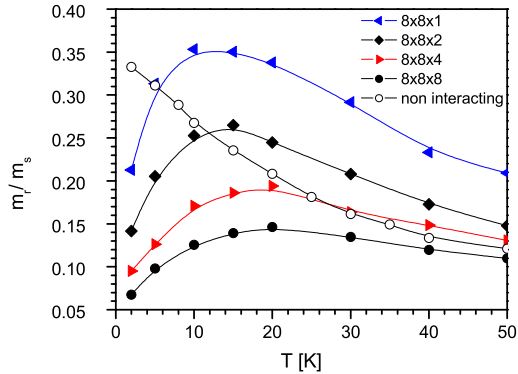}
\caption{Remanence as function of temperature for interacting systems with $M$ layers (full symbols), and for a noninteracting system (open symbols).
Each layer is composed by $8^2$ particles, placed on a square lattice. The interlayer and intralayer nearest neighbor distance  is $3.098 \AA$ The lines are guides to the eye. Error bars are smaller than the symbols.
\label{fig-sq1}}
\end{center}
\end{figure}

We start our calculations considering lattices with $M$ layers of $8^2$ particles and $a_x = a_y = a_z = 3.098 \AA$, as defined in Fig.~\ref{fig-redes}. These values of lattice parameter would correspond to a 20\% concentration in a simple cubic system.
 Fig.~\ref{fig-sq1} shows the thermal variation of the reduced remanence as a function of  $M$.
One can see from this figure that the remanence of interacting systems exhibits a maximum
at low temperatures, which is not present in the noninteracting particles
curve.  As shown in [17] for the $M=1$ system, the remanence peak separates distinct states of low magnetization. At low temperatures 
the absolute value of the dipolar energy is much larger than the thermal energy and an ordered state, due to the dipolar interaction, appears. In this case, magnetization patterns at $H=0$ show a very small contribution of $z$ magnetization and antiparallel alignment between lines normal to the previous field direction, the $y$-axis in this case. The energy required to satisfy the dipolar coupling between lines parallel to the $x$-axis would be too large and  an antiparallel alignment along the $y$-direction is favored. At the maximum, the two energies are closer, and thermal fluctuations provide enough energy to destroy the low temperature $y$ dipolar order. At higher temperatures, the thermal energy dominates, leading to a significant increase of the $z$ component of the magnetization, also resulting in a low remanence value.  This behavior is numerically well described by the average values of $|m_x|$, $|m_y|$ and $|m_z|$, defined as $\mu_x$, $\mu_y$ and $\mu_z$, respectively.  At $3\;$K, averaging over 10 samples, we have the values $\mu_x=503\mu_B$, $\mu_y=540\mu_B$ and $\mu_z=255\mu_B$, compatible with a mainly 
in-plane magnetization, with an important $y$ contribution due to the antiferromagnetic alignment. 
At the high temperature region, 
 for example at $70\;$K, the average values are $\mu_x=472\;\mu_B$,
$\mu_y=479\;\mu_B$ and $\mu_z=351\;\mu_B$, showing an increase of the out-of-plane component favored by thermal energy.

The remanence curves  for $M > 1$ are similar to the $M=1$ curve, with two low magnetization regions, one at low temperatures dominated by the dipolar coupling, in which antiparallel alignment in the directions normal to the applied field is present, and a high temperature disordered state with magnetic moments randomly aligned.
It is clear from Fig.~\ref{fig-sq1} that, as $M$ increases, for a given temperature the remanence decreases, and the peak slightly shifts to higher temperatures, being the interaction with out-of-plane neighbors the main cause of this behavior. 
Regarding the ordered low temperature state, the presence of layers below and above a given plane offers an alternative  direction for antiparallel alignment leading to a decrease of the remanence. Due to the long range character of the interaction, the dipolar energy per particle increases with system size up to a saturation value, therefore, it is natural to expect that increasing the  number of layers, the low temperature ordered state becomes more stable, reflecting in a shift of the peak position to higher temperatures. 

\begin{figure}
\begin{center}
\includegraphics{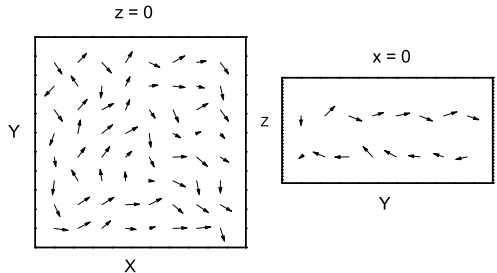}
\caption{Snapshots of the magnetic moments at surface layers of a $8\times 8\times 2$ system. 
\label{fig-snapq2}}
\end{center}
\end{figure}

In order to investigate the low temperature dependence on $M$, we have examined the magnetization patterns for  systems with $M=2$ and $M=8$. Figure~\ref{fig-snapq2} illustrates the behavior of the individual magnetic moments at $T=3\;$K,  for surface layers of a $M=2$ system. In this case, there are two alternative directions for the antiferromagnetic coupling, $y$ and $z$, but, as explained above, there is a predominance of the $y$ component since the number of neighbors in this direction is larger.  This can be confirmed  by looking at the average values of the magnetization components in the $x$,$y$ and $z$ directions. At $T = 3K$, considering 10 samples, the average values for the whole lattice are $\langle m_x\rangle = 403\mu_B $ and $\mu_x = 458\mu_B$, 
$\langle m_y\rangle = -9.47\mu_B$ and  $\mu_y = 504\mu_B$, $\langle m_z\rangle = 1.85\mu_B$ and  $\mu_z = 339\mu_B$. Clearly an antiparallel alignment on the $yz$ plane occurs, which is stronger along the $y$ direction.

For the $M=8$ system the  $y$ and $z$ are equivalent directions for the antiferromagnetic coupling.  The amount of  $z$ alignment increases at the expense  of  the $x$ contribution as can be seen from the average values for the lattice, $\langle m_x\rangle =246\mu_B$ and $\mu_x = 388\mu_B$, $\langle m_y\rangle = 14.3\mu_B$ and  $\mu_y = 517\mu_B$, $\langle m_z\rangle = -5.78\mu_B$ and  $\mu_z = 398\mu_B$. 

\begin{figure}
\begin{center}
\includegraphics{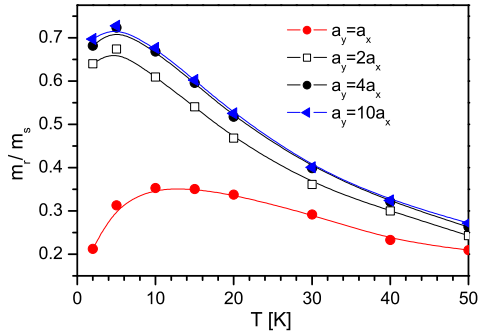}
\caption{Remanence as a function of temperature for different values of $a_y$ in $8\times 8\times 1$ lattices.
\label{fig-ay}}
\end{center}
\end{figure}

It is interesting to investigate the behavior of the remanence as the values of $a_y$ and $a_z$ are varied.  Figure~ \ref{fig-ay} shows the remanence curves for $M=1$ systems with $a_y = a_x, 2a_x$, $4a_x$ and $10a_x$. Since the antiparallel order in the $y$ direction is the main reason for the low value of remanence at low temperatures, as the $y$ distance is increased, the magnetic moments can follow the field more easily, and the remanence increases. 

\begin{figure}
\begin{center}
\includegraphics{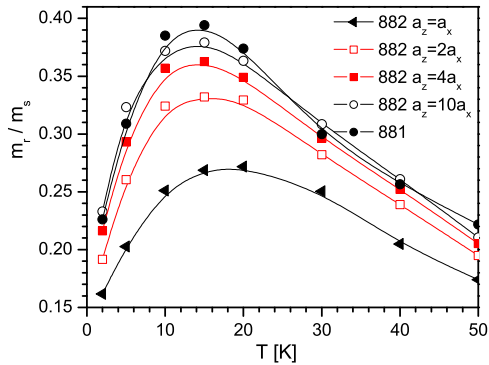}
\caption{Remanence as a function of temperature for different values of $a_z$ in $8\times 8\times 2$ lattices. The curve for a $8\times 8\times 1$ is also shown as a reference.
\label{fig-az}}
\end{center}
\end{figure}

The effect of varying $a_z$ is evident when we compare systems with $M=2$ and $a_z=a_x$, $2a_x$, $4a_x$ and $10a_x$ with a $M=1$ system. Figure~\ref{fig-az} illustrates the remanence curves for such systems, showing that, as the layers become more separated, the interlayer coupling decreases and the system approaches the $M=1$ behavior. 
\subsection{Triangular lattice}
\begin{figure}
\begin{center}
\includegraphics{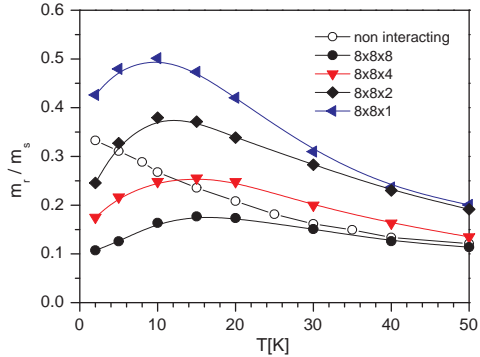}
\caption{Remanence as a function of temperature for  interacting systems with $M$ layers (full symbols), and for a non interacting system (open symbols). Each layer is composed by $8^2$ particles placed on a triangular lattice. 
\label{fig-tr1}}
\end{center}
\end{figure}

\begin{figure}
\begin{center}
\includegraphics{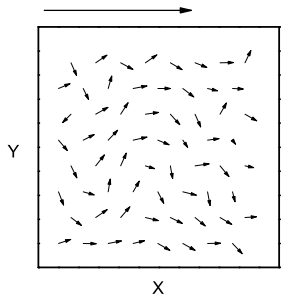}
\caption{Snapshot of the in-plane magnetization at $H = 0$ and $T=3$K, for a system of $8\times8\times1$ particles and triangular symmetry. The previous direction of the field is indicated by the arrow. 
\label{fig-tr2}}
\end{center}
\end{figure}

For layers with triangular symmetry, the number of in-plane neighbors is higher, so that any effects related to confinement of the magnetic moments in the $xy$ plane are enhanced.
Figure~\ref{fig-tr1} shows the remanence curves for triangular lattices with $a_t = 3.098 \AA$, obtained under the same conditions as the curves in Fig.\ref{fig-sq1}.
The remanence values are considerably larger than the ones in Fig. \ref{fig-sq1}, which is compatible with the picture of smaller $z$ component.
We also analyzed  the magnetization pattern for typical configurations of these triangular systems.
Figure \ref{fig-tr2} shows snapshots of the $x$ and $y$ components of the individual magnetic moments at $T = 3 K$ for $M = 1$. From this figure we can see that, at low temperature, the small value of the remanence is mainly caused by an antiferromagnetic ordering along the direction connecting the nearest neighbors clusters of the system. This effect, due to the
predominance of the dipolar interaction, is stronger than in the square lattice since the number of in-plane nearest neighbors, 6 in this lattice, is higher, as compared to 4 in the square lattice.  This behavior is numerically well described by the average values  of magnetization components at $T=3$K: $\langle m_x\rangle=571\mu_B$, $\langle m_y\rangle=-88.4\mu_B$, $\langle m_z\rangle=-25.9\mu_B$, $\mu_x =586\mu_B$, $\mu_y=433\mu_B$ and $\mu_z=285\mu_B$, compatible with a mainly  in-plane magnetization along the line joining nearest neighbors. At the high-temperature region, the system is disordered and the $z$ component of the magnetization increases, leading to a low remanence region. 

\section*{Discussion and conclusions}
The above results confirm the existence of two low-remanence yet distinct behaviors, one at low
temperature, where the dipolar energy dominates, and another, at higher
temperatures, where the thermal energy is responsible of the magnetic
disorder. 
The position and height of the peak separating those  regimes is related to the
strength of the dipolar
interactions, and depends on the lattice geometry. As known,
dipolar interactions favor closed circuit alignment, which may result in
antiparallel alignments between lines of parallel magnetic moments in rectangular lattices.
At low temperatures a small external magnetic field provides a prefered direction for the
parallel coupling, leading to two possibilities of antiparallel alignment in 3D systems.
On the other hand, for a 2D system, there is only one possible direction for the
antiparallel alignment. Figure \ref{fig-sq1} shows that for systems formed by sequential deposition of layers it is possible to
observe a transition from the 2D to the 3D behavior as the number of layers increases.
As an alternative direction for antiparallel alignment appears, the height of 
the remanence peak decreases reflecting the decrease of the magnetization along
the field direction.
Also, for a given number of layers, the variation of the distance between magnetic particles 
and  coordination number can drastically chance the effective 
dimensionality, as the dipolar interaction is enhanced along certain directions.
Figures  \ref{fig-ay} and  \ref{fig-az} show this dimensionality transition caused
by variation of the lattice parameter, and Figure \ref{fig-tr1} by the change in coordination number.
These aspects are responsible for difficulties in the analysis of the
hysteresis cycle, specially if one associates the area of the cycle to
the stability of the magnetization moment. Larger values of remanence are basically a consequence of a decrease in the
number of degrees of freedom, due to the confinement of the magnetic moment
to a plane or to a line.

 \section*{Acknowledgments}
In Chile the authors received financial support from FONDECYT under
grants \# 1010127, 7010127 and Millennium Science Nucleus
``Condensed Matter Physics'' P02-054F. In Brazil, the authors
acknowledge the support from FAPERJ, CNPq and Instituto de Nanoci\^encias/MCT. Paola R. Arias acknowledge MECESUP   support under grant USA0108.
%

\end{document}